\documentclass[ApJL, twocolumn]{aastex631}

\usepackage{graphicx}
\usepackage{amsmath}
\usepackage{color}
\usepackage[normalem]{ulem}
\usepackage{units}

\newcommand{\pistroke}{ \text{\protect\ooalign{\hidewidth\raisebox{-0.2ex}{--}\hidewidth\cr$\pi$\cr}}
}

\newcommand{\SPA}{School of Physics and Astronomy, Monash University, Clayton VIC 3800, Australia}
\newcommand{\OzGravMonash}{OzGrav: The ARC Centre of Excellence for Gravitational Wave Discovery, Clayton VIC 3800, Australia}

\shorttitle{Both black holes spin in merging binaries}
\shortauthors{Adamcewicz et al.}

\begin{document}

\title{Do both black holes spin in merging binaries? Evidence from GWTC-4 and astrophysical implications}

\author{Christian Adamcewicz}
\email{christian.adamcewicz@monash.edu}
\affiliation{\SPA}
\affiliation{\OzGravMonash}

\author{Nir Guttman}
\affiliation{\SPA}
\affiliation{\OzGravMonash}

\author{Paul D. Lasky}
\affiliation{\SPA}
\affiliation{\OzGravMonash}

\author{Eric Thrane}
\affiliation{\SPA}
\affiliation{\OzGravMonash}

\begin{abstract}
Angular momentum transport in high-mass stars is commonly modeled by extrapolating the behavior of better-observed low-mass stars.
According to the conventional picture, the cores of most black hole progenitors lose almost all of their angular momentum when their outer layers are ejected before core collapse.
Accordingly, \textit{most} black holes are expected to be born with dimensionless spin magnitudes of $\chi \lesssim 0.01$, even if \textit{some} black holes are born with non-negligible spin due to tidal interactions in a progenitor binary.
One might therefore expect to find a large fraction of $\chi\lesssim0.01$ black holes in merging binary black hole (BBH) systems.
We find that the conventional picture of angular momentum transport is in tension with data from LIGO--Virgo--KAGRA's fourth gravitational-wave transient catalog.
We find no support for a sub-population of BBH systems with $\chi\lesssim0.01$.
Neither do we find support for a sub-population with only one spinning black hole as expected for tidal spin-up scenarios.
Instead, we find evidence for two subpopulations in which \textit{both} black holes have non-negligible spin.
Approximately $84\%$ of BBH systems contain two black holes with modest spins $\chi\approx0.1$ and approximately $16\%$ contain two black holes with large spins $\chi\approx 0.8$.
These estimates come from our best-fit model, which is favored with natural log Bayes factors $\ln{\cal B}\gtrsim 3$ over models that require a sub-population of $\chi\lesssim 0.01$ black holes, and models that do not contain multiple spin sub-populations.
These results are difficult to reconcile with our current understanding of angular momentum transport.
\end{abstract}
 

\section{Introduction}\label{sec:introduction}

The release of the LIGO-Virgo-KAGRA collaboration's \citep[LVK's;][]{LIGOScientific:2014pky, VIRGO:2014yos, KAGRA:2020tym} fourth gravitational wave transient catalog \citep[GWTC-4;][]{GWTC4:intro, GWTC4:methods, GWTC4, GWTC4:data} brings the total number of detected binary black hole (BBH) mergers to 153.\footnote{
Where detections are thresholded on a false alarm rate of $< 1\,\mathrm{yr}^{-1}$, and a BBH is considered to be a merger with component masses inferred to be $\geq 3 \,M_\odot$ with 90\% credibility.
}
Made possible by improved detection capabilities \citep{Capote:2024rmo, LIGO:2024kkz, 2020PhRvD.102f2003B, LIGOO4Detector:2023wmz, membersoftheLIGOScientific:2024elc}, this expanded dataset provides the potential for new insights into the formation histories of these black holes, their progenitors and the binaries they constitute, many facets of which remain highly uncertain \citep[see reviews by][for example]{Callister:2024cdx, Spera:2022byb, Mandel:2018hfr, Mapelli:2018uds}.

One such area of uncertainty pertains to the spins of black holes in merging binaries.
Models for stellar spins are thought to be well calibrated via asteroseismology in low-mass stars \citep[e.g.,][]{Gehan:2018, Hermes:2017, Deheuvels:2015}.
However, a lack of such direct measurements leaves the spin properties of high-mass stars (including black hole progenitors) less certain.
\cite{Fuller:2019sxi} for instance \citep[see also,][]{Ma:2019cpr}, assume that angular momentum transport in black hole progenitors may be similarly efficient to that measured in low-mass stars.
Under this assumption, the authors show that BBH progenitors should lose the majority of their angular momentum via stripping of their outer layers before core collapse.
As such, \cite{Fuller:2019sxi} predict that black holes should typically be born with dimensionless spin magnitudes of $\chi \lesssim 0.01$.

A number of works have searched the previous GWTC-3 catalog \citep{KAGRA:2021vkt} for a subpopulation of such $\chi\lesssim0.01$ black holes, finding mixed, but ultimately inconclusive evidence \citep{Szemraj:2025fmm, Colloms:2025hib, Hussain:2024qzl, Adamcewicz:2023szp, Tong:2022iws, Mould:2022xeu, Callister:2022qwb, Galaudage:2021rkt, Kimball}.
Regardless, it is certain that a substantial fraction of the BBH population \citep[anywhere from 35\%-100\%;][]{Szemraj:2025fmm, Adamcewicz:2023szp, Tong:2022iws, Mould:2022xeu, Callister:2022qwb} merges with component spins inconsistent with $\chi\lesssim0.01$.
The reconstructed spin magnitude distribution peaks around $\chi \approx 0.2$ \citep[see][the first of which includes the most up-to-date inferences using GWTC-4]{GWTC4:rp, Hussain:2024qzl, Adamcewicz:2023szp, Callister:2023tgi, Edelman:2022ydv, Mould:2022xeu, Tong:2022iws, KAGRA:2021duu}.
These more rapidly rotating black holes may remain compatible with the theory put forth by \cite{Fuller:2019sxi} if black holes are spun up via tidal interactions in the progenitor binary.

In the tidal spin up scenario, one component of an isolated binary collapses to a black hole and raises tides on its Wolf-Rayet star companion \citep[a bare stellar core that has been stripped of its outer hydrogen layers;][]{Crowther:2006dd}.
As these tides dissipate, they produce a torque on the non-collapsed companion, giving the exposed stellar core a spin angular momentum that it retains after collapse \citep{Ma:2023nrf, Fuller:2022ysb, Hu:2022ubh, Olejak:2021iux, Bavera:2020inc, Belczynski:2017gds, Qin:2018vaa}.
This spun-up second-born component should typically be the less massive of the binary, but may be the more massive if the system has undergone mass-ratio reversal \citep[see][for example]{Broekgaarden:2022nst, Zevin:2022wrw, Olejak:2021iux}.

Tidal interactions may also, in some cases, spin up both BBH progenitors during isolated evolution.
This process involves two progenitors with very similar masses and a low initial orbital separation becoming tidally locked early in life.
This results in the spin up of both components, which in turn become chemically homogeneous, subduing their radial expansion along with the occurrence of mass and angular momentum loss \citep{Mandel:2018hfr, Marchant:2016wow, Mandel:2015qlu}.
However, the chemically homogeneously evolved binaries are expected to make up only a small fraction of the total binary black hole population.

Another possible scenario is that high-spin black holes are created through hierarchical mergers.
In dense stellar environments, it may be possible for two black holes to merge into one, with the product of this merger (referred to as a second-generation black hole) subsequently being captured in orbit with a yet another black hole \citep{Li:2022gly, Doctor:2021qfn, Doctor:2019ruh, Mapelli:2020xeq, Rodriguez:2019huv, Gerosa:2017kvu, Fishbach:2017dwv,Kimball,gwtc2_hierarchical, Mahapatra:2021hme, Mahapatra:2022ngs, Mahapatra:2024qsy}.
Since second-generation black holes inherit the orbital angular momentum of the predecessors binary, they are expected to have spin magnitudes $\chi \sim 0.7$ \citep{Tichy:2008du}, although, see \cite{Borchers:2025sid} and \cite{McKernan:2023xio} for exceptions.
A second-generation black hole may merge with a first generation-black hole, meaning only one component in the binary has a high spin magnitude, or it may merge with another second-generation black hole, meaning both components are rapidly rotating.
The latter scenario is expected to be rare \citep{Li:2022gly, Gerosa:2019zmo, Rodriguez:2019huv, Mahapatra:2022ngs}.

Finally, black holes may be spun up through accretion via material from its binary companion in an isolated system \citep[e.g.,][]{Olejak:2021iux, Bavera:2020uch, vanSon:2020zbk}, or even by interstellar gas in active galactic nuclei \citep[AGN;][]{McKernan:2023xio, Bogdanovic:2007hp}.
Accretion between components in isolated binaries is commonly invoked to explain X-ray binary black holes' rapid spins \citep[e.g.,][]{Shao:2020tin, Qin:2018sxk, Podsiadlowski:2003}.
However, X-ray binaries probably do not contribute significantly to the population of merging BBH systems \citep{Mandel:2018hfr, Gallegos-Garcia:2022rve, Liotine:2022vwq}.

In this work, we search for subpopulations of black holes with different spin characteristics using data from  GWTC-4.
In particular, we look for a subpopulation of systems with $\chi \lesssim 0.01$ as predicted by \cite{Fuller:2019sxi}.
The remainder of this manuscript is structured as follows.
In Section~\ref{sec:methods}, we describe the statistical framework used to probe for multiple spin subpopulations in the gravitational-wave data.
In Section~\ref{sec:zero_spins}, we employ this framework under the assumption that non-spun-up black holes merge with negligibly small spin magnitudes $\chi \lesssim 0.01$.
In Section~\ref{sec:nonzero_spins}, we repeat this analysis under the assumption that non-spun-up black holes have spin magnitudes that are not negligible, $\chi \gg 0$.
Finally, in Section~\ref{sec:discussion}, we discuss the astrophysical implications of our findings along with potential follow-ups to the study presented here.

\section{Framework}\label{sec:methods}
We model the distribution of black hole spin with a four-component mixture model.
The primary black hole spin $\chi_1$ can be ``low-spin,'' drawn from the distribution $\pi_0(\chi|\Lambda)$; or it can be ``high-spin,'' drawn from the distribution $\pi_s(\chi|\Lambda)$ (The $s$ subscript stands for ``spun up'').
The secondary black hole spin $\chi_2$ can be drawn from the same two distributions.
The distributions $\pi_0(\chi|\Lambda)$ and $\pi_s(\chi|\Lambda)$ are conditioned on a set of hyperparameters $\Lambda$, which control their shape.

With two possible distributions for both the primary and secondary black hole, there are four permutations: neither black hole is spun up, only the primary black hole is spun up, only the secondary black hole is spun up, or both black holes are spun up.
Our mixture model is therefore:
\begin{align}\label{eq:full_spin_model}
    \pi(\chi_1,\chi_2|\Lambda) &= 
    \lambda_0 \pi_0(\chi_1|\Lambda) \pi_0(\chi_2|\Lambda) \nonumber \\
    &+ \lambda_1 \pi_s(\chi_1|\Lambda) \pi_0(\chi_2|\Lambda) \nonumber \\
    &+ \lambda_2 \pi_0(\chi_1|\Lambda) \pi_s(\chi_2|\Lambda) \nonumber \\
    &+ \lambda_b \pi_s(\chi_1|\Lambda) \pi_s(\chi_2|\Lambda).
\end{align}
The coefficients $\{\lambda_0, \lambda_1, \lambda_2, \lambda_b\}$ are mixing fractions, which add to unity:
\begin{align}
    \lambda_0 + \lambda_1 + \lambda_2 +
    \lambda_b = 1 .
\end{align}

We further suppose that the high-spin population is related to the low-spin population as follows.
The $\chi$ value of high-spin black holes is given by the sum of two random variables: 
\begin{align}
    \chi_s = \chi_0 + \chi_+ .
\end{align}
Here, $\chi_0$ is a low-$\chi$ value drawn from $\pi_0(\chi|\Lambda)$ and $\chi_+$ is a spin-up value distributed according to a truncated Gaussian:
\begin{equation}\label{eq:spin_up_model}
    \pi_+(\chi_+|\chi_0,\Lambda) \propto 
    \mathcal{N}(\chi_+|\mu_+,\sigma_+) H(\chi_+) H(1- \chi_0 - \chi_+).
\end{equation}
Here, $\mathcal{N}(x|\mu,\sigma)$ is a normal distribution, where the mean $\mu_+$ and width $\sigma_+$ are a subset of hyperparameters $\Lambda$.
Meanwhile, $H(x)$ is a Heaviside step function, employed so that $\chi_s$ is on the interval $(\chi_0, 1)$.
It follows that  
\begin{align}
    \pi_s(\chi_s | \Lambda) = \int d\chi_0 \,
    \pi_0(\chi_0 | \Lambda) \,
    \pi_+(\chi_s - \chi_0| \chi_0, \Lambda) .
\end{align}
We leave the distribution $\pi_0(\chi|\Lambda)$ to be defined under two different assumptions in Sections~\ref{sec:zero_spins}~and~\ref{sec:nonzero_spins}.

We assume that the distribution of black hole spin magnitude factorizes from the distributions of mass, spin tilt, and distance.
We model the primary mass $m_1$ and mass ratio $q=m_1/m_2$ using the preferred broken-power-law + two-peak model from \cite{GWTC4:rp}.
We also fit the distribution of redshifts using a power-law model \citep{Fishbach:2018edt, LIGOScientific:2020kqk, KAGRA:2021duu, GWTC4:rp}, and the distribution of cosine spin tilts using a Gaussian + isotropic mixture model \citep{Talbot:2017yur, LIGOScientific:2020kqk, KAGRA:2021duu, GWTC4:rp}.
We use the priors detailed in Appendix~\ref{app:priors} for the spin magnitude model hyperparameters, and the same priors as \cite{GWTC4:rp} for the mass, tilt, and redshift model hyperparameters.

In analyzing the population, we use the 153 BBH mergers detected with a false-alarm-rate (FAR) $< \unit[1]{yr^{-1}}$ in GWTC-4 \citep{GWTC4} \citep[this follows the threshold set by][]{GWTC4:rp}.
The source properties of these gravitational-wave events up to and including the O3b observing run are fit using the \texttt{IMRPhenomXPHM} waveform model \citep{Pratten:2020ceb}.
Events from O4a are mostly fit using the \texttt{NRSur7dq} waveform model \citep{Varma:2019csw}, with the exception of events that lie in regions of parameter space outside of this approximant’s range of validity, in which case we revert to \texttt{IMRPhenomXPHM}.
These data products, taken from official LVK releases \citep{GWTC4, GWTC4:data, pe_data, pe_data_3, pe_data_2}, are supplemented with additional samples that assume, \textit{a priori}, component spins of $\chi=0$ (more on this in Section~\ref{sec:zero_spins}).
Apart from the modified priors, these additional samples are produced under identical assumptions to the official LVK data products \citep{GWTC4:methods, GWTC4:data, pe_data, pe_data_3, pe_data_2}.
We account for selection effects in the population using the ``found injections'' method \citep{Tiwari:2017ndi, Mandel:2018mve, Farr:2019}, with the official LVK sensitivity estimates \citep{Essick:2025zed, sens_data}.
Hierarchical inference of the population hyperparameters $\Lambda$ is performed with the nested sampler \textsc{Dynesty} \citep{Speagle:2020} and a modified version of \textsc{GWPopulation} \citep{Talbot:2024yqw, Talbot:2019okv}, which utilises the Bayesian inference library \textsc{BILBY} \citep{Ashton:2018jfp, Romero-Shaw:2020owr}.
For more on hierarchical inference in gravitational wave astronomy, see \cite{Thrane:2019}.

\section{Assuming typical spins are negligibly small}\label{sec:zero_spins}
We first consider the case in which black holes are born with very small dimensionless spin magnitudes $\chi_0 \lesssim 0.01$ as predicted by \cite{Fuller:2019sxi}.

\subsection{Model}\label{subsec:zero_spins_model}

To model these small spin magnitudes, we approximate spins of $\chi_0 \lesssim 0.01$ as $\chi_0 = 0$.
For all intents and purposes, these values are indistinguishable.
However, assuming $\chi=0$, we can use dedicated posterior samples for each gravitational-wave event that are calculated assuming \textit{a priori} that the black holes have zero spin.
This prevents the undersampling that can arise when one employs an initially broad prior and then uses importance sampling with a very narrow target distribution.
We expand on this framework in \cite{Adamcewicz:2023szp} (see, in particular, Appendix A), and in Section~\ref{subsec:zero_spins_results} below.
The distribution of $\chi_0$ is then a delta function,
\begin{equation}\label{eq:zero_spin_model}
    \pi_0(\chi_0|\Lambda) = \delta(\chi_0).
\end{equation}
The spin of the spun-up black hole is
$\chi_s = \chi_+$, which is distributed according to Eq.~\ref{eq:spin_up_model},
\begin{equation}\label{eq:zero_final_spin_model}
    \pi_s(\chi_s|\Lambda) = \pi_+(\chi_s|\chi_0=0,\Lambda).
\end{equation}
The two distributions given in Eqs.~\ref{eq:zero_spin_model}~and~\ref{eq:zero_final_spin_model} are inserted into Eq.~\ref{eq:full_spin_model} to give the final spin model for the BBH population.
This model is similar to those used in \cite{Galaudage:2021rkt}, \cite{Tong:2022iws} and \cite{Adamcewicz:2023szp}, except for the fact that spinning black holes in this work are truncated-Gaussian distributed as opposed to Beta distributed.

We do not account for spin-based selection effects in these analyses.
Firstly, the data products required for this would be costly to produce.
Secondly, spin magnitudes, particularly in the case of small spins like the ones being probed in this model, have been shown to produce a negligible observation bias \citep{Ng:2018neg, Golomb:2022bon, Adamcewicz:2023szp, Essick:2025zed, Szemraj:2025fmm}.

\subsection{Results}\label{subsec:zero_spins_results}

\begin{figure*}
    \centering
    \includegraphics[width=0.8\textwidth]{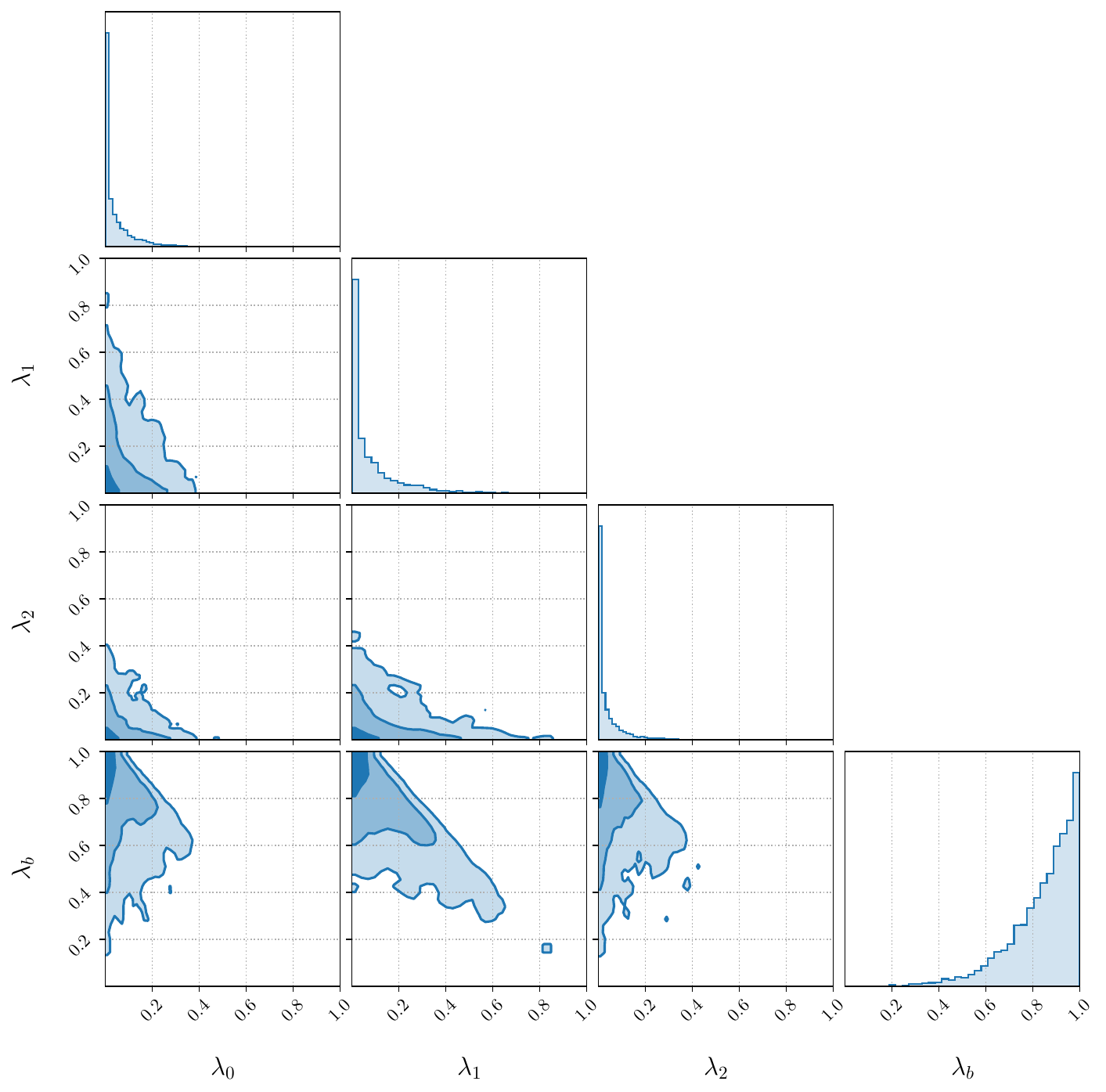}
    \caption{
    Posterior distributions for population branching fractions assuming non-spun-up black holes have spin magnitudes of $\chi_0=0$.
    The fractions $\lambda_0$, $\lambda_1$, $\lambda_2$ and $\lambda_b$ represent subpopulations of binaries in which neither black hole is spun up, the primary black hole is spun up, the secondary black hole is spun up, and both black holes are spun up, respectively.
    From darkest to lightest, contours in two-dimensional panels give the 50\%, 90\% and 99\% credible regions of the posterior.
    We see strong support for $\lambda_b \approx 1$, implying that almost all black holes in merging binaries have some measurable spin magnitude.
    This may indicate that nearly all black holes are spun up, or that it is incorrect to assume that black holes are typically born with negligibly small spins in the absence of a spin-up mechanism.
    }
    \label{fig:z_lambda_corner}
\end{figure*}

Using the above model, we obtain posteriors for the branching fractions to each of the four subpopulations described in Eq.~\ref{eq:full_spin_model}.
These are plotted in Fig.~\ref{fig:z_lambda_corner}.
We see that the posterior peaks at $\lambda_b \approx 1$ and $\lambda_0 \approx \lambda_1 \approx \lambda_2 \approx 0$, implying that the best fit to the data is a model in which all black holes have a non-zero spin magnitude.
There is no support for a sub-population where $\chi_1=0$ as one would expect in the conventional tidal spin-up framework.
There is no support for a sub-population where $\chi_2=0$ as one would expect in tidal spin-up scenarios with mass-ratio reversal~\citep{Broekgaarden:2022nst}.

Our above conclusion is supported by comparing the evidence for models with different subsets of allowed subpopulations.
A model with only spinning black holes ($\lambda_b = 1$) is mildly favored over a model that allows for a subpopulation of binaries with non-spinning black holes ($\lambda_0 > 0$, $\lambda_b > 0$; $\ln \mathcal{B} = 1.1$), and a model that allows for non-spinning binaries as well as binaries in which one or both components are spun up ($\lambda_0 > 0$, $\lambda_1 > 0$, $\lambda_2 > 0$, $\lambda_b > 0$; $\ln \mathcal{B} = 3.7$).
Furthermore, the assumption that only one black hole may be spinning in a given binary ($\lambda_1 > 0$, $\lambda_2 > 0$, $\lambda_0 \geq 0$, $\lambda_b=0$) is strongly disfavored ($\ln \mathcal{B} \gtrsim 10$).
These results are summarized in Table~\ref{tab:evidence}.

From the above results, we see that the existence of subpopulations containing non-spinning black holes is statistically disfavored.
However, binaries in which one or both black holes are effectively non-rotating cannot be definitively ruled out; $\lesssim 35\%$ (90\% credibility) of systems may have one or two negligibly small component spins.

\begin{figure*}
    \centering
    \includegraphics[width=0.85\textwidth]{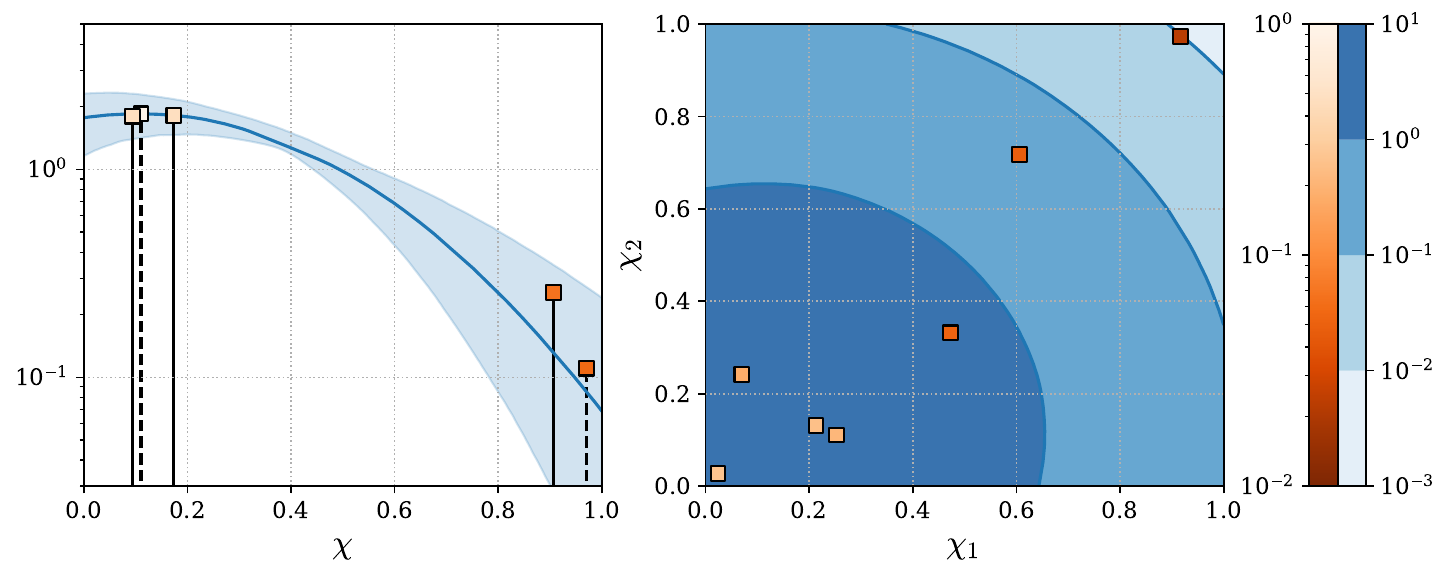}
    \caption{
    Population predictive distribution for the best-fit model when assuming black holes are typically born with $\chi_0=0$ (blue).
    The left panel gives the marginal distribution for one component ($\chi_1$ and $\chi_2$ are identically distributed in this model), while the right panel gives the joint spin magnitude distribution.
    Both panels give probability in logarithmic scale.
    In this best-fit model, $\lambda_b=1$ implying all black holes in the population are spun-up.
    This results in a Gaussian distribution of spin magnitudes.
    Over-plotted in orange is the $\pistroke$ fit to GWTC-4 from \cite{pistroke}.
    In the one-dimensional plot, the solid and dashed tails give the $\pistroke$ fit for $\chi_1$ and $\chi_2$ respectively.
    The heights of the points are scaled such that the tallest point coincides with the median fit to the population model.
    The color-bar gives the normalized weight.
    }
    \label{fig:ppd_default}
\end{figure*}

Taking the favored model from this subset (one in which all black holes spin $\lambda_b=1$), we plot the population predictive distribution for BBH spin magnitudes in Fig.~\ref{fig:ppd_default}.
In spite of a lack of support for a separate non-spinning subpopulation, we see strong support as $\chi \rightarrow 0$, with a broad tail extending out to $\chi \approx 1$.
This particular model implies identically and independently distributed spin magnitudes between the two binary components, and is functionally identical to the default spin magnitude model presented in \cite{GWTC4:rp}.
In Fig.~\ref{fig:ppd_default}, we also over-plot the $\pistroke$ result from \cite{pistroke}.
Introduced in \cite{Payne:2022xan}, the $\pistroke$ formalism identifies a weighted sum of delta functions that are the maximum population likelihood solution (for the space of all possible population models).\footnote{In the limit that the number of events goes to infinity, the $\pistroke$ distribution becomes the true astrophysical distribution.}
The locations and heights of these delta functions are a representation of the data, and highlight features that models might try to fit.
We see that the fit to the population model under-plotted in blue roughly recovers the features in the $\pistroke$ analysis, but perhaps over-predicts the density in regions of (high-$\chi_1$, low-$\chi_2$) and (low-$\chi_1$, high-$\chi_2$).

\begin{figure}
    \centering
    \includegraphics[width=\columnwidth]{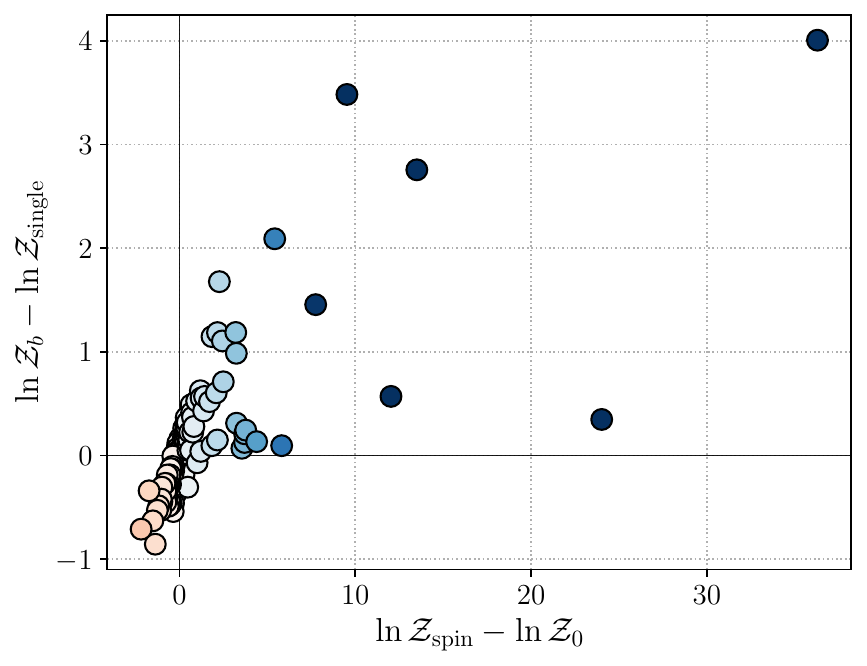}
    \caption{
    Comparison of spin scenarios for the 153 BBH events in GWTC-4.
    The horizontal axis gives the natural log Bayes factor comparing the hypothesis that at least one component in the binary has a non-zero spin magnitude, to the hypothesis that both components have spin magnitudes of zero.
    The colour of each marker is also determined by the same natural log Bayes factor; this helps guide the eye to the events that are most clearly spinning.
    The vertical axis gives the natural log Bayes factor comparing the hypothesis that both components in the binary spin to the hypothesis that only one component spins.
    Events tend to cluster around zero -- presenting no evidence in any direction.
    As events show more evidence for component spins of any sort, they consistently show more evidence for both components in the binary having a non-zero spin magnitude, as opposed to only one.
    }
    \label{fig:spin_evidence}
\end{figure}

To obtain data products required for these analyses, we perform Bayesian inference on each event in the catalog three times, providing four sets of samples when combined with the official LVK data products \citep{GWTC4, GWTC4:data, pe_data, pe_data_3, pe_data_2}.
Taking the official parameter estimation in which both black holes spin $(\chi_1>0, \chi_2>0)$, we reproduce the samples assuming only the primary spins $(\chi_1>0, \chi_2=0)$, then assuming only the secondary spins $(\chi_1=0, \chi_2>0)$, and finally assuming both black holes are non-rotating $(\chi_1=0, \chi_2=0)$~\citep[see][for more information]{Adamcewicz:2023szp}.
As such, we obtain Bayesian evidences for these four scenarios for each BBH merger in the catalog, which we denote $\mathcal{Z}_b$, $\mathcal{Z}_1$, $\mathcal{Z}_2$ and $\mathcal{Z}_0$, respectively.
We then find the evidence for any non-zero component spins in a given binary (coming from the primary, secondary, or both components) as
\begin{equation}
    \mathcal{Z}_\mathrm{spin} =
    \frac{1}{3}\left( \mathcal{Z}_b + \mathcal{Z}_1 + \mathcal{Z}_2 \right).
\end{equation}
This is equivalent to the Bayesian evidence obtained via inference when weighting all three spin hypotheses equally in the prior.
Similarly, we obtain the evidence for the hypothesis that only one black hole spins (either the primary or secondary) in a given binary as
\begin{equation}
    \mathcal{Z}_\mathrm{single} =
    \frac{1}{2}\left( \mathcal{Z}_1 + \mathcal{Z}_2 \right).
\end{equation}

We use these evidences in Fig.~\ref{fig:spin_evidence} to compare the support for the several different spin hypotheses for each event in GWTC-4.
For most events, we do not find compelling evidence for any one spin hypothesis over another.
Meanwhile, a small subset of events show relatively strong evidence for some kind of non-zero component spin.
These events also tend to exhibit a moderate amount of evidence supporting both black holes in the binary spinning as opposed to only one.
These findings match with those presented in the population analyses above.

\section{Assuming typical spins are measurably non-zero}\label{sec:nonzero_spins}

Next, we consider the case in which black holes are typically born with spin magnitudes higher than those predicted in \cite{Fuller:2019sxi}, such that they are no longer negligibly small.
In doing so, we let the distribution of $\chi_0$ vary with the data.

\subsection{Model}\label{subsec:nonzero_spins_model}
We model $\chi_0$ as a Gaussian distributed variable, truncated over physical values of $\chi_0=0$ to $\chi_0=1$:
\begin{equation}\label{eq:nonzero_spin_model}
    \pi_0(\chi_0|\Lambda) \propto 
    \mathcal{N}(\chi_0|\mu_0,\sigma_0) H(\chi_0) H(\chi_0 - 1).
\end{equation}
Again, the mean $\mu_0$ and width $\sigma_0$ are hyperparameters $\Lambda$ to be fit to the data.\footnote{
\label{footnote:model_freedom}
We also experiment by allowing $\mu_+$ and $\sigma_+$ to be different for the singly spun-up subpopulations, and both spun-up subpopulations.
We do not find a meaningful difference in our results with this added model freedom.
}

Contrary to the model used in Section~\ref{sec:zero_spins}, accounting for spin-based selection effects using this framework does not require any additional inputs.
As such, spin-based selection effects are incorporated into the analyses in this section.
We find that excluding spin selection effects (as was done in Section~\ref{sec:zero_spins}) underestimates the uncertainty on the spin hyperparameters governing the high-spin subpoulations.
However, the peaks of the posteriors do not shift meaningfully.

\subsection{Results}\label{subsec:nonzero_spins_results}

\begin{figure*}
    \centering
    \includegraphics[width=0.8\textwidth]{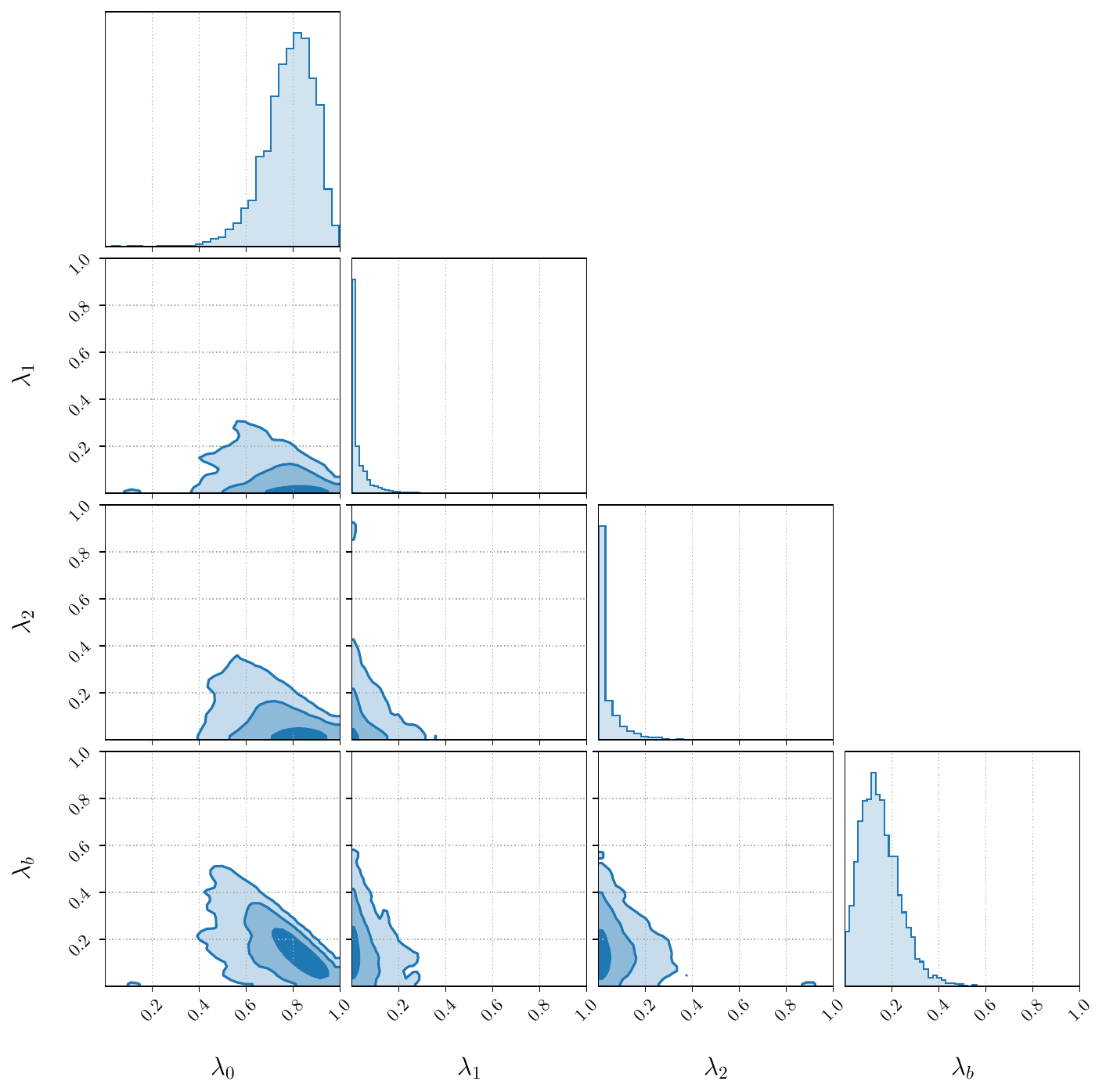}
    \caption{
    Posterior distributions for population branching fractions assuming non-spun-up black holes have non-zero, Gaussian distributed spin magnitudes.
    The fractions $\lambda_0$, $\lambda_1$, $\lambda_2$ and $\lambda_b$ represent subpopulations of binaries in which neither black hole is spun up, the primary black hole is spun up, the secondary black hole is spun up, and both black holes are spun up beyond their initial Gaussian distributed spin, respectively.
    From darkest to lightest, contours in two-dimensional panels give the 50\%, 90\% and 99\% credible regions of the posterior.
    We find a peak at $\lambda_0 \approx 0.8$, indicating that most, but likely not all black holes follow some Gaussian like distribution of small spin magnitudes.
    Meanwhile, $\lambda_1$ and $\lambda_2$ peak at zero, indicating that systems in which only one black hole is spun up are rare.
    On the other hand, $\lambda_b$ peaks at $\approx 0.1$, meaning the data favors a considerable fraction of binaries in which both components are spun up beyond spins typical of the rest of the population.
    }
    \label{fig:nz_lambda_corner}
\end{figure*}

The branching fractions for each subpopulation are plotted in Fig.~\ref{fig:nz_lambda_corner}.
We observe peaks at $\lambda_0 = 0.80_{-0.21}^{+0.13}$ and $\lambda_b = 0.14_{-0.11}^{+0.16}$ (90\% credibility), while $\lambda_1$ and $\lambda_2$ peak at zero.
This indicates strong support for a subpopulation in which neither black hole is spun up beyond the Gaussian $\chi_0$ distribution, a second subpopulation in which both black holes are spun up, \textit{but no subpopulations in which only one black hole is spun up}.

A similar narrative is made clear by comparing evidence values for models that only allow for certain subsets of the four possible subpopulations.
We find that the best-fit model is one in which either neither black hole is spun up, or both black holes are spun up, but never just one: $\lambda_0>0$, $\lambda_b>0$, $\lambda_1=0$,  $\lambda_2=0$.
This is preferred over a model in which neither black hole is spun up beyond the Gaussian $\chi_0$ distribution \citep[$\lambda_0=1$; functionally equivalent to the best-fit $\lambda_b=1$ model from Section~\ref{sec:zero_spins}, and the default spin model from][]{GWTC4:rp} by a natural log Bayes factor of $\ln \mathcal{B} = 3.6$.
Relative to the best-fit model, including subpopulations in which only one black hole is spun up is disfavored ($\lambda_0 > 0$, $\lambda_1 > 0$, $\lambda_2 > 0$, $\lambda_b > 0$; $\ln \mathcal{B} = 1.2$).
Similarly, a model that only allows for one black hole to be spun up and not both is disfavored ($\lambda_0 > 0$, $\lambda_1 > 0$, $\lambda_2 > 0$, $\lambda_b = 0$; $\ln \mathcal{B} = 4.6$).
These quantities are summarized in Table~\ref{tab:evidence}.

\begin{figure*}
    \centering
    \includegraphics[width=0.85\textwidth]{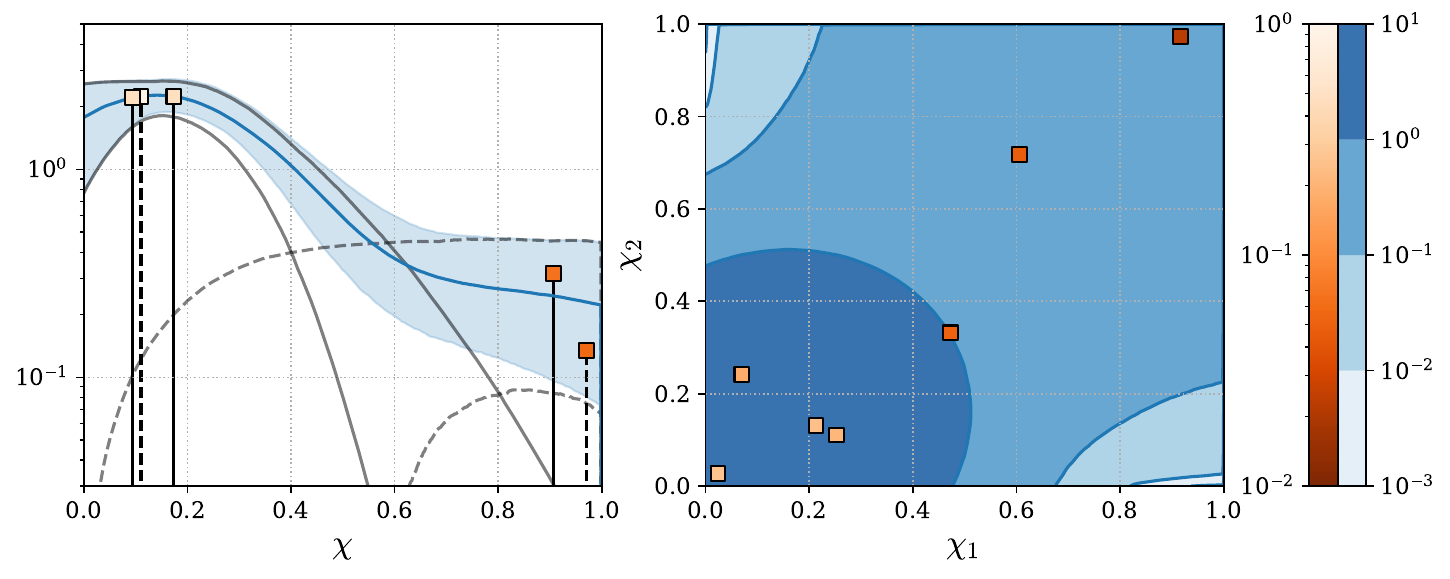}
    \caption{
    Population predictive distribution for the best-fit model when assuming black holes are typically born with non-zero, Gaussian-distributed spins.
    This is the best-fit model of all considered in this paper.
    In blue, the left panel gives the marginal distribution for one component ($\chi_1$ and $\chi_2$ are identically distributed), while the right panel gives the joint spin magnitude distribution.
    Both panels are in logarithmic scale.
    In this best-fit model, only $\lambda_0$ and $\lambda_b$ are allowed to be non-zero, implying that for any given system, neither black hole is spun up, or both black holes are spun up.
    The 90\% credible contributions from these two subpopulations are over-plotted in solid and dashed gray lines respectively.
    Over-plotted in orange, is the $\pistroke$ fit to GWTC-4 from \cite{pistroke}.
    In the one-dimensional plot, the solid and dashed tails give the $\pistroke$ fit for $\chi_1$ and $\chi_2$ respectively.
    The heights of the points are scaled such that the tallest point coincides with the median fit to the population model.
    The color-bar gives the normalized weight.
    }
    \label{fig:ppd_spin-up}
\end{figure*}

In Fig.~\ref{fig:ppd_spin-up}, we plot the population predictive distribution for the best fit model.
We see the distribution peaks near $\chi \approx 0.1$, with a long tail out to $\chi=1$.
Having a second subpopulation of binaries in which both black holes are spun up produces a positive correlation in $(\chi_1, \chi_2)$, where merger-rates drop off in regions where only one black hole is rapidly rotating.
While spin magnitudes are not independently distributed in this model, they are distributed identically in $\chi_1$ and $\chi_2$, due to the lack of subpopulations in which only one black hole is spun up.
Again, we over-plot the $\pistroke$ result from \cite{pistroke}.
We see that the population model better matches with the \cite{pistroke} results than the model from Section~\ref{sec:zero_spins}, predicting higher densities out to (high-$\chi_1$, high-$\chi_2$), and lower densities at (high-$\chi_1$, low-$\chi_2$) and (low-$\chi_1$, high-$\chi_2$).

\begin{figure*}
    \centering
    \includegraphics[width=0.8\textwidth]{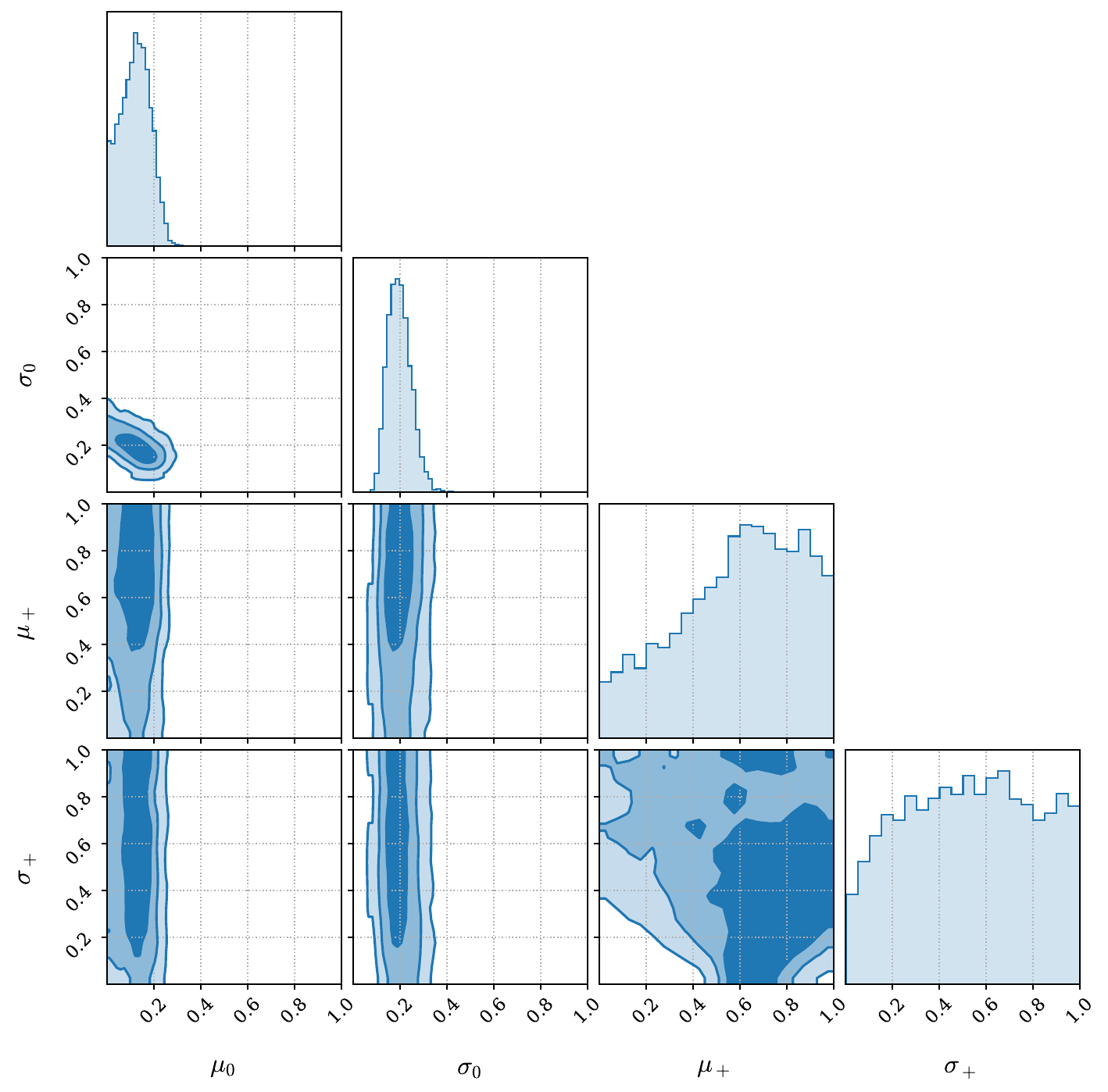}
    \caption{
    Posterior for the hyperparameters governing the shapes of the subpopulations' spin magnitude distribution.
    From left to right, we have the mean $\mu_0$ and width $\sigma_0$ of the default non-spun-up spin distribution, and the mean $\mu_+$ and width $\sigma_+$ of the distribution of spins gained via spin-up.
    From darkest to lightest, contours in two-dimensional panels give the 50\%, 90\% and 99\% credible regions of the posterior.
    This plot gives the posteriors from an analysis allowing for all four subpopulations, however, the posteriors are nearly identical to the favored model which disallows binaries in which only one black hole is spun up.
    }
    \label{fig:mu_sigma_post}
\end{figure*}

In Fig.~\ref{fig:mu_sigma_post}, we plot the posterior distributions for the hyperparameters governing the shapes of the spin subpopulations.
The low- (non-spun-up-) spin distribution has a confidently inferred peak at $\mu_0 = 0.12_{-0.10}^{+0.10}$ and width of $\sigma_0 = 0.19_{-0.07}^{+0.09}$.
Meanwhile, the distribution of spins gained via spin up peaks at $\mu_+ = 0.63_{-0.51}^{+0.33}$ and has a width of $\sigma_+ = 0.53_{-0.45}^{+0.42}$.
The latter two hyperparameters are less constrained than the former, showing support across most of the prior range.
However, we see a subtle preference for spins gained from spin-up to be consistently large -- preferring large peak values of $\mu_+$ and smaller variances via $\sigma_+$.
Heuristically, smaller values of $\sigma_+$ lead to a bimodality in the distribution; as $\sigma_+ \rightarrow 0$, the spun-up subpopulation becomes a translation of of the non-spun-up subpopulation by a value of $\mu_+$.
Conversely, larger values of $\sigma_+$ lead to the spun-up subpopulation being a smeared or positively skewed version of the non-spun-up subpopulation, implying less or no bimodality in the population.
The former scenario appears to be mildly preferred.

\begin{table*}
    \centering
    \begin{tabular}{|c|c c c c|c c|}
        \hline
        $\chi_0$ & $\lambda_0$ & $\lambda_1$ & $\lambda_2$ & $\lambda_b$ & $\ln \mathcal{B}$ & $\Delta \ln \mathcal{L}_{\max}$ \\
        \hline
        \hline
        $0$ & $1$ & $0$ & $0$ & $0$ & $-153.7$ & $-157.3$ \\
        \hline
        $0$ & $0$ & $0$ & $0$ & $1$ & $0.0$ & $0.0$ \\
        \hline
        $0$ & $0.02^{+0.16}_{-0.02}$ & $0$ & $0$ & $0.98^{+0.02}_{-0.16}$ & $-1.1$ & $+0.7$ \\
        \hline
        $0$ & $0$ & $0.98^{+0.02}_{-0.22}$ & $0.02^{+0.22}_{-0.02}$ & $0$ & $-10.5$ & $-9.9$ \\
        \hline
        $0$ & $0.02^{+0.18}_{-0.02}$ & $0.92^{+0.08}_{-0.26}$ & $0.02^{+0.23}_{-0.02}$ & $0$ & $-12.0$ & $-10.9$ \\
        \hline
        $0$ & $0.01^{+0.16}_{-0.01}$ & $0.03^{+0.29}_{-0.03}$ & $0.01^{+0.15}_{-0.01}$ & $0.88^{+0.11}_{-0.31}$ & $-3.7$ & $0.0$ \\
        \hline
        \hline
        $>0$ & $1$ & $0$ & $0$ & $0$ & $0.0$ $[0.0]$ & $0.0$ $[0.0]$ \\
        \hline
        $>0$ & $0$ & $0$ & $0$ & $1$ & $-5.0$ $[-3.1]$ & $0.0$ $[+0.2]$ \\
        \hline
        $>0$ & $0.84^{+0.11}_{-0.16}$ & $0$ & $0$ & $0.16^{+0.16}_{-0.11}$ & $+6.8$ $[+3.6]$ & $+10.4$ $[+3.8]$ \\
        \hline
        $>0$ & $0$ & $0.23^{+0.75}_{-0.23}$ & $0.77^{+0.23}_{-0.75}$ & $0$ & $-4.1$ $[-3.4]$ & $+0.6$ $[+0.2]$ \\
        \hline
        $>0$ & $0.86^{+0.13}_{-0.30}$ & $0.06^{+0.24}_{-0.06}$ & $0.03^{+0.25}_{-0.03}$ & $0$ & $-0.4$ $[-1.0]$ & $+0.8$ $[+0.5]$ \\
        \hline
        $>0$ & $0.80^{+0.13}_{-0.21}$ & $0.01^{+0.11}_{-0.01}$ & $0.01^{+0.15}_{-0.01}$ & $0.14^{+0.16}_{-0.11}$ & $+3.2$ $[+2.4]$ & $+9.0$ $[+4.6]$ \\
        \hline
    \end{tabular}
    \caption{
    Natural log Bayes factors $\ln \mathcal{B}$ and differences in maximum likelihoods $\Delta \ln \mathcal{L}_{\max}$ for each model fit to the gravitational-wave data.
    Where available, Bayes factors and likelihood values in square brackets include spin selection effects, while those outside brackets do not.
    Models are distinguished by whether they assume non-spun-up black holes have negligibly small spin magnitudes ($\chi_0=0$) or non-negligible spin magnitudes ($\chi_0>0$), as well as which subpopulations they allow for.
    Here, $\lambda_0$, $\lambda_1$, $\lambda_2$, and $\lambda_b$ give the fraction of systems in which neither black hole is spun up, the primary black hole is spun up, the secondary black hole is spun up, and both black holes are spun up, respectively.
    In models where the fraction is fit to the data, the quoted value gives the median and 90\% credible intervals.
    Bayes factors and relative maximum likelihoods are quoted relative to the default Gaussian spin magnitude model from \cite{GWTC4:rp}.
    This model is functionally identical to a model with $\chi_0=0$ and $\lambda_b=1$ as well as a model with $\chi_0>0$ and $\lambda_0=1$.
    }
    \label{tab:evidence}
\end{table*}

\section{Discussion}\label{sec:discussion}
We begin our discussion with results from Section~\ref{sec:zero_spins}.
In line with a number of works from GWTC-3, we do not find evidence for a subpopulation of BBH components with negligible spin magnitudes \citep{Szemraj:2025fmm, Hussain:2024qzl, Callister:2022qwb, Mould:2022xeu}.
Although, see \cite{Colloms:2025hib}, who find that the GWTC-3 population is consistent with natal spins of $\chi \approx 0.04$.
Separately, we highlight the studies in \cite{Tong:2022iws} and \cite{Adamcewicz:2023szp}.
While the results of these works cannot rule out the hypothesis that all merging binary black holes have measurable spin magnitudes, they are unique in that they find that a subpopulation of non-spinning black holes ($\approx 30\%$ of all binaries) is modestly favored by the data.

In producing this manuscript, we revisited these analyses to find that a normalization error in the population likelihood resulted in an overestimate in the evidence for a non-spinning subpopulation.
We elaborate more on this and detail the updated conclusions from \cite{Tong:2022iws} and \cite{Adamcewicz:2023szp} with this error corrected in Appendix~\ref{app:gwtc3_revision}.
While these corrected results remain inconclusive as to whether there exists a subpopulation of non-spinning black holes, they now modestly favor the hypothesis that no-such subpopulation exists, in line with other works \citep{Hussain:2024qzl, Callister:2022qwb, Mould:2022xeu}.
The conclusions from \cite{Adamcewicz:2023szp} -- that there is not a substantial subpopulation of binaries in which only the secondary black hole spins, and that systems with measurable spins show more support for both black holes spinning as opposed to only the primary spinning -- remain unchanged.

If future studies can conclude that a significant subpopulation of black holes with negligibly small spins does not exist, one of two things must be true.
The first is that the assumption that black hole progenitors lose nearly all of their angular momentum before core collapse is incorrect.
If this is true, by extension, it may imply that models for stellar rotation tuned to low-mass stars are not applicable to high-mass stars \citep{Fuller:2019sxi, Ma:2019cpr}.
The alternative is that virtually all merging binary black holes (or progenitors) are spun-up via a process akin to one of those discussed in Section~\ref{sec:introduction}.

Assuming the former to be true, we now turn our attention to the analyses presented in Section~\ref{sec:nonzero_spins}.
If we take the models presented here at face-value, we find that black holes, absent some additional spin-up process, merge with typical spins of $\chi_0 \lesssim 0.3$ (based on the median inferred values of $\mu_0$ and $\sigma_0$).
Meanwhile, $\approx 14 \%$ of the population undergoes some process by which both black holes are spun-up, adding typical values of $\chi_+ \approx 0.1{-}1.0$ (again based on median inferred values of $\mu_+$ and $\sigma_+$) on top of their typical natal spins $\chi_0$.
A seemingly negligible fraction of systems undergo a similar spin-up process in only the primary component ($\lesssim 8 \%$) or only the secondary component ($\lesssim 11 \%$).

This qualitative structure of two subpopulations in $(\chi_1,\chi_2)$ -- one with both black hole spin magnitudes being small, and another with both being large -- matches the GWTC-3 findings of \cite{Hussain:2024qzl}, where the authors fit the BBH spin magnitude population to various mixture models that broadly consist of two covariant Gaussian distributions.
Furthermore, we find substantial overlap in the branching fractions between these two subpopulations, with high-spin binaries making up $16_{-11}^{+16}\%$ of the population in this work (when assuming no singly-spun-up subpopulations), and $20_{-18}^{+18}\%$ in \cite{Hussain:2024qzl}.

It is somewhat difficult to square this tendency for both black holes to be spun up with standard theories.
As discussed in Section~\ref{sec:introduction}, one possibility is this subpopulation is a result of BBH systems being formed via the chemically homogeneous channel, in which two isolated progenitors are spun up after becoming tidally locked.
However, this mechanism typically requires that the binary in question has a roughly equal mass ratio, which is in conflict with the evidence for more rapidly spinning binaries typically having less-equal mass ratios \citep{GWTC4:rp, Heinzel:2023hlb, Adamcewicz:2023mov, Adamcewicz:2022hce, KAGRA:2021duu, Callister:2021fpo}.
Furthermore, it seems unlikely that such a subpopulation is produced via hierarchical mergers.
This is due to the lack of binaries with a single rapidly-spinning component implying that first-generation -- second-generation mergers are exceptionally rare.
Meanwhile, the relative abundance of systems with both components rapidly spinning would indicate that mergers between two second-generation black holes are common, in contradiction with the expectation that such mergers should be rare \citep{Mahapatra:2022ngs, Li:2022gly, Gerosa:2019zmo, Rodriguez:2019huv}.

An alternative explanation may be that both black holes are spun up via accretion of interstellar material, perhaps in AGN disks \citep{McKernan:2023xio, Bogdanovic:2007hp}.
This mechanism could also help to explain the anisotropy (conflicting with typical dynamical formation scenarios), and the peak away from orbital alignment \citep[conflicting with typical isolated formation scenarios;][]{Yu:2020iqj, Farr:2017uvj, Stevenson:2017dlk, Talbot:2017yur, Rodriguez:2016vmx, Vitale:2015tea} in the BBH cosine tilt distribution inferred with GWTC-4 \citep{GWTC4:rp}.
This is because the accreted gas can produce a preferred spin orientation for black holes in alignment with the AGN disk (breaking isotropy) that is not necessarily in alignment with the BBH orbits themselves \citep{McKernan:2023xio}.
That said, black hole dynamics in AGN is relatively poorly understood, and so we hesitate to invoke them every time we observe a feature in the BBH population that we do not understand.

In light of the above discussion, it is not clear if the approach and results presented in Section~\ref{sec:nonzero_spins} are hinting at distinct subpopulations of BBH mergers with different spin properties, or a single smoothly varying population in which $\chi_1$ and $\chi_2$ are positively correlated.
Further checking for bimodality as opposed to unimodality in the distribution of $(\chi_1, \chi_2)$ could be interesting in this regard.
Furthermore, one might test to see if the transition between the modeled spin subpopulations is accompanied by sharp transitions in the distributions of other parameters (e.g., mass, mass ratio or spin tilts), or by smoother correlations.
The former of these would be more clearly indicative of distinct subpopulations with more disparate formation histories.
\cite{Pierra:2024fbl} for example, use GWTC-3 data to compare the hypothesis that the BBH spin distribution exhibits a smooth correlation with mass, to the hypothesis that the spin distribution changes sharply at a specific transition mass, finding the latter is mildly preferred \citep[see also][]{Li:2025rhu, Antonini:2024het, Guo:2024wwv, Li:2023yyt, Godfrey:2023oxb, Wang:2022gnx, Mould:2022ccw}.

We also note that distinct high- and low-spin subpopulations do not necessarily correspond to subpopulations of spun-up and non-spun-up black holes.
While this is perhaps the most natural explanation for these features, one cannot necessarily exclude the possibility of bimodality in the natal black hole spin distribution.

We cannot exclude the possibility that distinct, smaller subpopulations are being lost within the slowly-spinning and rapidly-spinning subpopulations uncovered here.
\cite{Hussain:2024qzl} for example, find that the slowly-spinning subpopulation exhibits a small preference for an anti-correlation in $(\chi_1,\chi_2)$.
Such a feature could be a result of a fraction of binaries within this subpopulation undergoing tidal spin up of the secondary component, while others have tidally spun-up primaries.
This small increase in either components spin magnitude from tidal spin-up would be consistent with predictions from \cite{Ma:2023nrf}.
However, we would expect that separate parameterizations of the spin-up distribution for binaries with both black holes spun up and the spin-up distribution for binaries with only one black hole spun up would recover such a feature if it existed in GWTC-4.
We do not find this to be the case (see Footnote~\ref{footnote:model_freedom}).

We reiterate that we cannot definitively rule out the hypothesis that a minority of binaries ($\lesssim 35\%$; disfavored by $\ln \mathcal{B} = 3.7$ relative to the default Gaussian spin model) contain one or more approximately non-rotating black holes.
If we insist that some merging black holes are not spinning, we can limit the branching fraction to different scenarios: only the primary spinning ($\lesssim 23\%$), only the secondary spinning ($\lesssim 11\%$), and neither black hole spinning ($\lesssim 13\%$).
These findings are roughly in-line with those of \cite{Adamcewicz:2023szp}, however, we are now confidently able to rule out the hypothesis that only one black hole can be spinning in any given binary ($\ln \mathcal{B} \gtrsim 10$).
This implies that, under the assumptions provided, neither tidal spin up, or first-generation -- second-generation hierarchical mergers can be the root-cause for a majority of the non-negligible spins in the BBH population.
From the high-spin subpopulation uncovered in Section~\ref{sec:nonzero_spins}, these scenarios do not appear to account for the highest-spin BBH systems in the population either.

In lieu of the above discussion and speculation, it is possible that the results presented here are not representative of the true distribution of BBH spins.
\cite{Miller:2024sui} for one, show that inferences on BBH spin distributions can be inaccurate for reasons that are difficult to diagnose -- even in simulated scenarios where the true, injected distribution is known exactly.
In Appendix~\ref{app:result_checks}, we perform a number of checks on our analyses to search for signs of model misspecification and systematic error.
Namely, we investigate individual events that drive the inferred spin distribution and check for simple degeneracies in the distributions of spin magnitude, tilt, and mass ratio.
We do not find any causes for concern from these checks.

\section*{Acknowledgements}
We thank Hui Tong and Sharan Banagiri for helpful discussions which motivated the validation studies conducted in Appendix~\ref{app:result_checks}.
We also thank Sylvia Biscoveanu for providing a helpful review during internal LVK circulation.

We acknowledge support from the Australian Research Council (ARC) Centre of Excellence CE230100016, LE210100002, and ARC DP230103088.
This material is based upon work supported by NSF's LIGO Laboratory which is a major facility fully funded by the National Science Foundation.
The authors are grateful for computational resources provided by the LIGO Laboratory and supported by National Science Foundation Grants PHY-0757058 and PHY-0823459.

\appendix

\section{Prior details}\label{app:priors}

In Table~\ref{tab:priors}, we list the priors used for the spin-magnitude hyperparameters $\Lambda$ during hierarchical inference (see Eq.~\ref{eq:full_spin_model}, as well as Eq.~\ref{eq:spin_up_model}, Eq.~\ref{eq:zero_spin_model}-\ref{eq:zero_final_spin_model} and Eq.~\ref{eq:nonzero_spin_model}).
Priors used on hyperparameters governing the mass, redshift and spin tilt distributions are identical to those listed in \cite{GWTC4:rp}.

For the branching fractions $\lambda_0$, $\lambda_1$, $\lambda_2$ and $\lambda_b$, we use a Dirichlet prior that equally weights the four subpopulations, while keeping them normalized (such that the sum of the branching fractions equals one).
While Dirichlet priors with a shape parameter of $\alpha=1$ are a common choice due to their uniformity in higher dimensions, the marginal prior on each branching fraction in this case heavily disfavors values of $\lambda \approx 1$.
In many of the models tested in this work, likelihoods are maximized as $\lambda \rightarrow 1$ (as a single subpopulation dominates).
As a result, we find that a shape parameter of $\alpha = 1$ sporadically causes issues during inference due to under-sampling in regions of hyperparameter space where the population likelihood is maximized.
Through experimentation, we find that a shape parameter of $\alpha=\frac{1}{3}$ is a good compromise, providing ample support across values of $\lambda \approx 0$ to $\lambda \approx 1$ and resolving these sampling issues.
From experimentation with different values of $\alpha$, we find that this choice does not meaningfully impact our results.
However, this choice may slightly reduce the Occam penalty suffered as a result of adding additional subpopulations to a model.

\begin{table}
    \centering
    \begin{tabular}{|c c l|}
        \hline
        Hyperparameter & Prior & Description \\
        \hline
        \hline
        $\mu_+$ & $\mathcal{U}(0, 1)$ & Mean of the Gaussian spin-up distribution.\\
        \hline
        $\sigma_+$ & $\mathcal{U}(0, 1)$ & Width of the Gaussian spin-up distribution.\\
        \hline
        $\mu_0$ & $\mathcal{U}(0, 1)$ & Mean of the Gaussian initial-spin distribution.\\
        \hline
        $\sigma_0$ & $\mathcal{U}(0, 1)$ & Width of the Gaussian initial-spin distribution.\\
        \hline
        $\lambda_0$ & $\mathrm{Dir}(\frac{1}{3}|\lambda_1,\lambda_2,\lambda_b)$ or $0$ & Fraction of systems in the non-spun-up subpopulation.\\
        \hline
        $\lambda_1$ & $\mathrm{Dir}(\frac{1}{3}|\lambda_0,\lambda_2,\lambda_b)$ or $0$ & Fraction of systems in the spun-up primary subpopulation.\\
        \hline
        $\lambda_2$ & $\mathrm{Dir}(\frac{1}{3}|\lambda_0,\lambda_1,\lambda_b)$ or $0$ & Fraction of systems in the spun-up secondary subpopulation.\\
        \hline
        $\lambda_b$ & $\mathrm{Dir}(\frac{1}{3}|\lambda_0,\lambda_1,\lambda_2)$ or $0$ & Fraction of systems in the both-spun-up subpopulation.\\
        \hline
    \end{tabular}
    \caption{
    List of priors used for hyperparameters $\Lambda$ governing the spin magnitude model.
    We also include a brief description of each hyperparameter.
    Here, $\mathcal{U}(a,b)$ denotes a uniform distribution from $a$ to $b$, and $\mathrm{Dir}(\alpha|\lambda)$ is a Dirichlet prior with a shape parameter $\alpha$, dependent on other branching fractions $\lambda$.
    In some model variations, a subset of branching fractions $\lambda$ are set to zero.
    Of course, for any given sample, the branching fractions $\lambda_i$ sum to unity.
    }
    \label{tab:priors}
\end{table}

\section{Revisiting zero-spin subpopulations in GWTC-3}\label{app:gwtc3_revision}

In this Section, we elaborate on the normalization error that affected the results of \cite{Tong:2022iws} and \cite{Adamcewicz:2023szp}.
We also present updated versions of these analyses, still using GWTC-3 data, but with this error corrected.

As explained in Appendix A of \cite{Adamcewicz:2023szp}, when fitting the data to subpopulations with $\chi = 0$, each subpopulation requires a unique set of posterior samples for every event in which sampling is performed dependent on one of the following four assumptions: $(\chi_1=0,\chi_2>0)$, $(\chi_1>0,\chi_2=0)$, $(\chi_1=0,\chi_2=0)$, or $(\chi_1>0,\chi_2>0)$.
Of course, the samples for each subpopulation also require unique weights, where the numerator is the proposed model for the given subpopulation, and the denominator is the uninformative prior assumed during the fiducial posterior sampling \citep[again, see Appendix A of][]{Adamcewicz:2023szp}.

Previous versions of \textsc{GWPopulation} \citep{Talbot:2024yqw} computed the fiducial sampling prior for redshift on a numerically normalized grid, then interpolated along the grid to obtain the prior probability for each sample.
Unbeknownst to the authors of \cite{Adamcewicz:2023szp} and \cite{Tong:2022iws}, the maximum redshift of this grid (thus the overall normalization of the fiducial redshift prior) is, in some instances, determined by the largest redshift posterior sample in the dataset.
Typically, this should not bias population results, as this method introduces a small, constant normalization error that is independent of the population model and its hyperparameters, thus factorizes with the population likelihood.
However, the edits made to \textsc{GWPopulation} for the analyses in \cite{Adamcewicz:2023szp} and \cite{Tong:2022iws} made it so that this gridded redshift prior calculation was repeated for each subpopulation.
As each subpopulation relied on different batches of posterior samples (thus had unique maximum redshifts), the normalization error described above could vary between subpopulations, which biased inferences on the branching fractions between them.

In this work, we base our code on an updated version of \textsc{GWPopulation} that no longer has this issue.
Below, we provide versions of the GWTC-3 results presented in \cite{Adamcewicz:2023szp} and \cite{Tong:2022iws} with this error corrected.

\subsection{Adamcewicz et al. (2024)}

\begin{figure}
    \centering
    \includegraphics[width=0.45\textwidth]{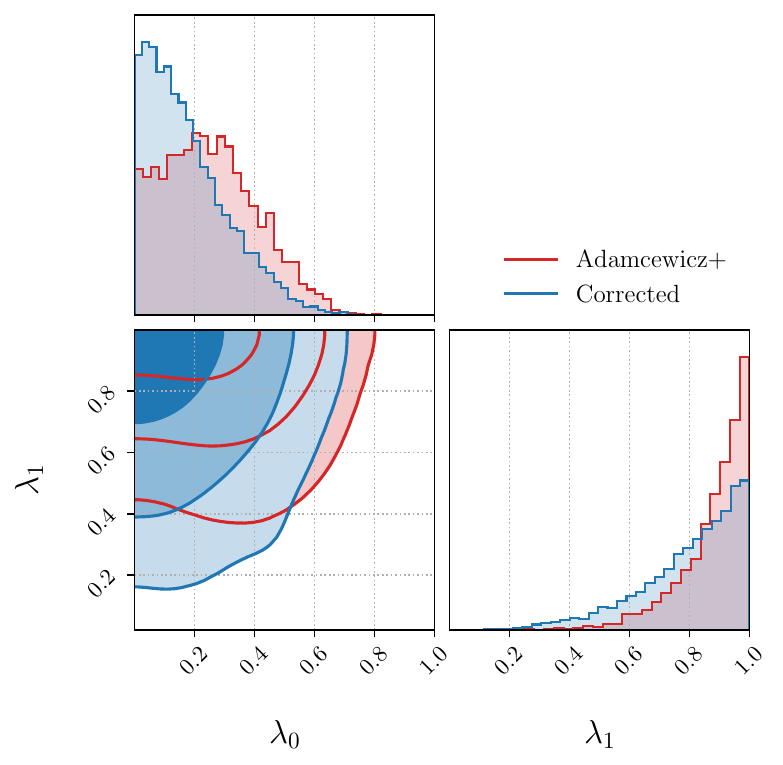}
    \caption{
    Updated Figure~1 from \cite{Adamcewicz:2023szp} with the normalization error corrected (blue), compared to the original results (red).
    Unlike the corresponding Figure in \cite{Adamcewicz:2023szp}, both sets of results here include GW191109\_010717 and GW200129\_065458.
    Posterior corner plot for the fraction of BBH systems with negligible spin $\lambda_0$, and the fraction of spinning BBH systems with $\chi_1 > 0$ as opposed to $\chi_2 > 0$, $\lambda_1$.
    From darkest to lightest, shading in the two-dimensional panel gives the 50\%, 90\%, and 99\% credible intervals.
    The posterior for $\lambda_1$ remains effectively unchanged, while the posterior for $\lambda_0$ is shifted towards zero.
    }
    \label{fig:Adamcewicz2024_fig1}
\end{figure}

We begin with the key conclusions from \cite{Adamcewicz:2023szp}.
In Figure~\ref{fig:Adamcewicz2024_fig1}, we re-plot the posterior distributions for the branching fractions assuming all binaries contain either no spinning black holes, or one spinning black hole.
See Figure~1 of \cite{Adamcewicz:2023szp} and supporting text for more information.
The inferred fraction of systems with the primary mass black hole spinning as opposed to the secondary mass black hole remains about the same, with the primary being the one spinning in at least $55\%$ of binaries, with peak support when $100\%$ of spinning black holes are the primaries in their given binaries.
Support for a zero-spin subpopulation has decreased, with such systems making up, at most, $34\%$ of the population.
However, the single-spin framework is now disfavored relative to the both-spin framework by an even larger natural log Bayes factor of $\ln \mathcal{B} = 5.9$ (difference in maximum natural log likelihood of $\Delta \ln \mathcal{L}_{\max} = 5.5$).
A model including both-spin and single-spin subpopulations is statistically indistinguishable from a simpler model that excludes single-spin subpopulations ($\ln \mathcal{B} = 0.6$ in favor of the simpler both-spin model).

\subsection{Tong et al. (2022)}

\begin{figure}
    \centering
    \includegraphics[width=0.75\textwidth]{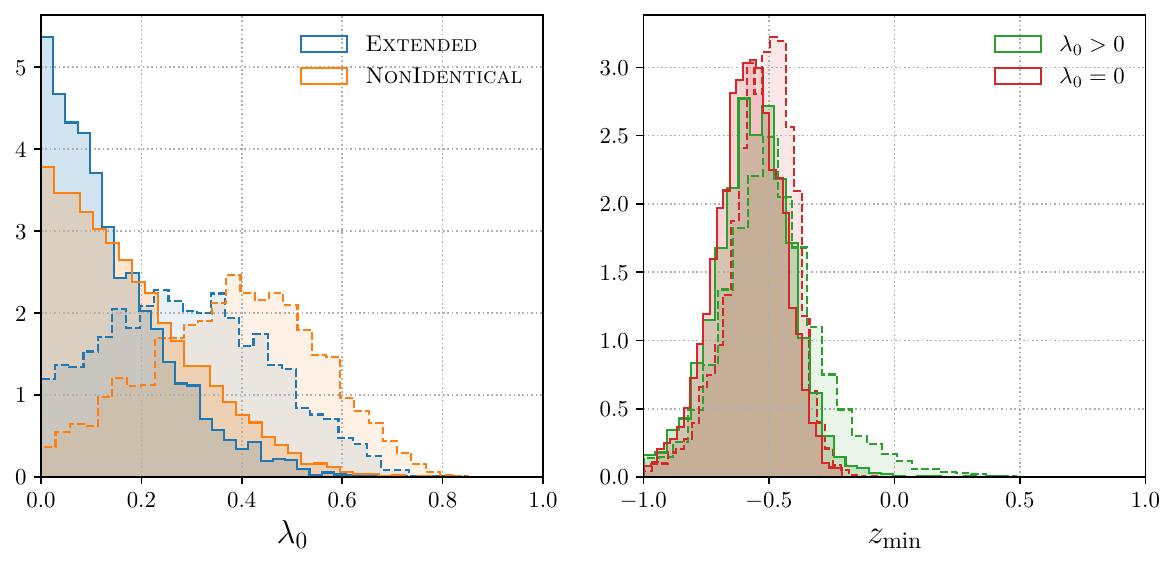}
    \caption{
    Updated Figure~1 from \cite{Tong:2022iws} with the normalization error corrected (solid lines), compared to original \cite{Tong:2022iws} results (dashed lines).
    Note that these plots do not exclude GW191109\_010717.
    The left panel gives the posterior of $\lambda_0$, the fraction of binaries in the non-spinning subpopulation.
    The blue trace assumes $\chi_1$ and $\chi_2$ are identically distributed, while the orange trace assumes they are not.
    As in \cite{Tong:2022iws}, the \textsc{IsoSubPop} variations of these models produce nearly identical posteriors on $\lambda_0$.
    Both models now peak as $\lambda_0 \rightarrow 1$.
    The right panel gives the posterior on the minimum allowed cosine spin tilt.
    The green trace is the posterior when a subpopulation of non-spinning black holes is allowed, while the red trace shows the posterior when this subpopulation is not allowed.
    The posterior with $\lambda_0>0$ is now more consistent with the $\lambda_0=0$ result than it was in \cite{Tong:2022iws}.
    We do not include a comparison with GWTC-2 in this plot.
    }
    \label{fig:Tong2022_fig1}
\end{figure}

We now provide updates from the analyses in \cite{Tong:2022iws}.
We plot the posteriors of $\lambda_0$ (the fraction of systems in the zero-spin subpopulation) and $z_{\min}$ (the minimum allowed cosine spin tilt) in Figure~\ref{fig:Tong2022_fig1}.
See Figure~1 of \cite{Tong:2022iws} and surrounding text for more information.
We see that there is less support for a non-spinning subpopulation than reported in \cite{Tong:2022iws}.
The inferred minimum allowed tilt remains consistent with the results from \cite{Tong:2022iws} when a non-spinning subpopulation is disallowed.
The inclusion of a non-spinning subpopulation no longer significantly affects the inferred value of $z_{\min}$.
We also include an updated version of Table III from \cite{Tong:2022iws}, which compares the evidence for the different models explored in \cite{Tong:2022iws}, in Table~\ref{tab:Tong2022_table3}.
There are a number of takeaways from these updated results.
A subpopulation of non-spinning BBH systems is no longer preferred.
The inclusion of a cut-off in cosine spin tilt ($z_{\min}$) is still preferred by a natural log Bayes factor of $\ln \mathcal{B} \approx 2$.
Non-identical distributions of $\chi_1$ and $\chi_2$ are generally disfavored.

\begin{table}
    \centering
    \begin{tabular}{|c|c c|c c|}
        \hline
        Model & $\ln \mathcal{B}$ & $\Delta \ln \mathcal{L}_{\max}$ & $\chi_1, \chi_2$ identical? & binaries with $z < z_{\min}$ \\
        \hline
        \hline
        \textsc{NonIdentical} & -3.08 & -0.57 & no & none \\
        \textsc{Extended} & -2.67 & -1.46 & yes & none \\
        \textsc{IsoSubPop} & -3.41 & -1.11 & yes & dynamical-like \\
        \textsc{NonIdentical IsoSubPop} & -4.49 & -0.96 & no & dynamical-like \\
        \hline
        \textsc{NonIdentical} with $\lambda_0=0$ & -0.94 & 0.00 & no & none \\
        \textsc{Extended} with $\lambda_0=0$ & 0.0 & -0.55 & yes & none \\
        \textsc{Extended} with $z_{\min}=-1$ & -4.42 & -2.08 & yes & none \\
        \hline
        \textsc{Default} & -2.62 & -0.57 & yes & yes \\
        \hline
    \end{tabular}
    \caption{
    Updated Table~III from \cite{Tong:2022iws}, providing relative evidences for the models explored in \cite{Tong:2022iws}.
    These results include GW191109\_010717.
    }
    \label{tab:Tong2022_table3}
\end{table}

\section{Validation}\label{app:result_checks}

In this section, we perform supplementary analyses to help validate our findings.

\subsection{Mass ratio and tilt dependencies}

The effective inspiral spin
\begin{equation}\label{eq:chieff}
    \chi_\mathrm{eff} = \frac{\chi_1 \cos t_1 + q \chi_2 \cos t_2}{1 + q},
\end{equation}
is a commonly studied parameter due to being approximately constant throughout the binary inspiral, as well as being typically well constrained relative to other BBH spin parameters \citep{Damour:2001tu}.
Given the latter point, it seems plausible that misspecifying the distributions of mass ratio $q$, or (cosine) spin tilts $\{\cos t_1, \cos t_2\}$, could lead to biased inferences on the component spin magnitudes $\{\chi_1,\chi_2\}$.

In Section~\ref{sec:nonzero_spins}, for example, forcing the low- and high-spin subpopulations to have identical mass ratio and spin tilt distributions may be problematic.
Given that the low-spin subpopulation appears to make up $\approx 80\%$ of the population, one would expect the mass ratio distribution to be predominantly informed by these low-spin binaries.
Even though high-spin binaries may tend to have more unequal mass ratios than the rest of the population \citep[e.g.,][]{Adamcewicz:2023mov, KAGRA:2021duu}, the relatively strong preference for equal mass ratios inferred from the low-spin subpopulation may be extrapolated to high-spin binaries due to a lack of model flexibility.
Combining this potentially spurious preference for $q \sim 1$ with the expression for $\chi_\mathrm{eff}$ in Eq.~\ref{eq:chieff}, we can see how inferences on spin magnitudes may be biased.
A similar line of reasoning can be laid out for spin tilts.

\begin{figure}
    \centering
    \includegraphics[width=0.8\textwidth]{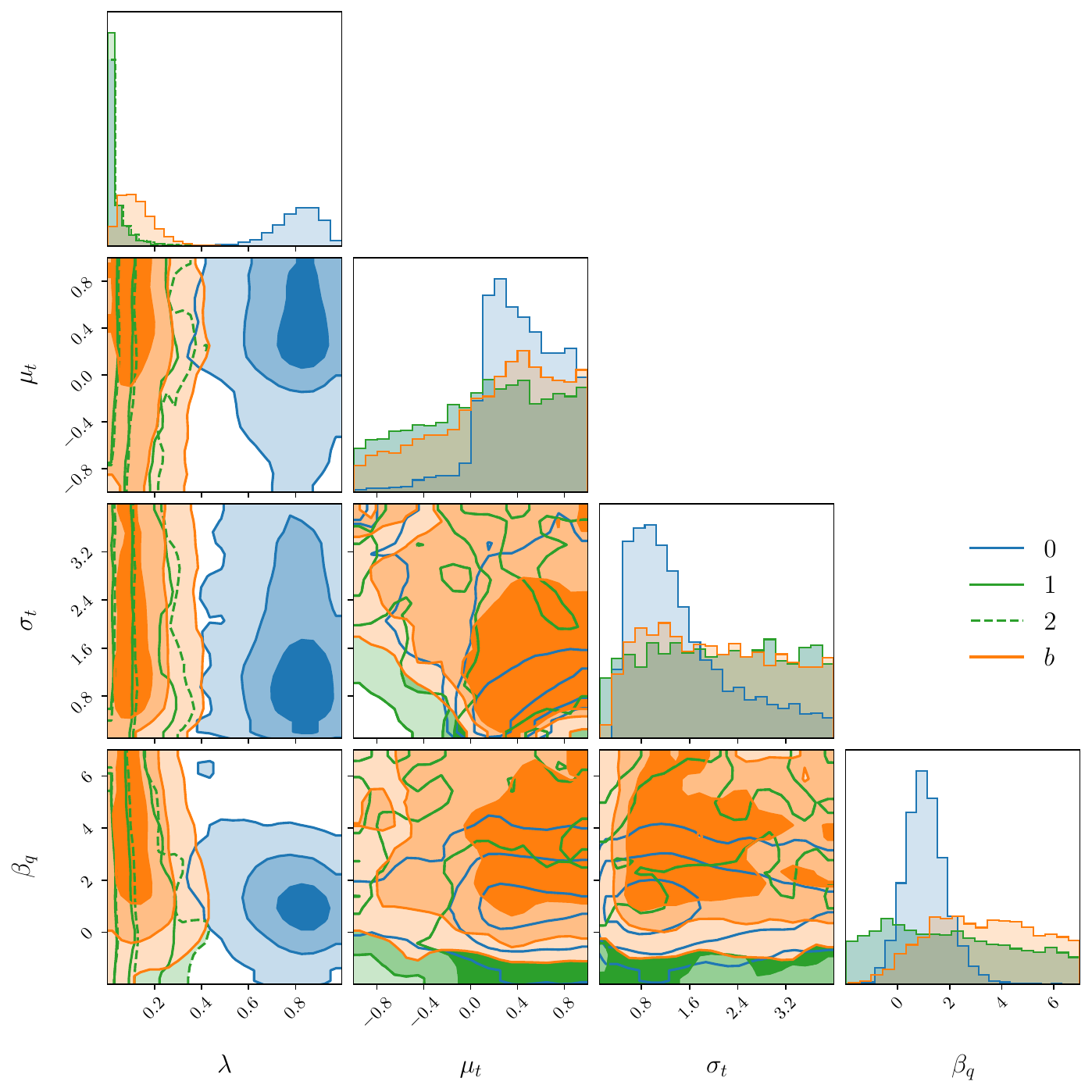}
    \caption{
    Posterior distributions for the mixing fraction $\lambda$, mean and width of the cosine tilt distribution $\mu_t$ and $\sigma_t$, and power-law index of the mass ratio distribution $\beta_q$ for each subpopulation.
    The posteriors in blue, solid-green, dashed-green and orange give the posteriors for the non-spun-up, primary-spun-up, secondary-spun-up, and both-spun-up subpopulations respectively.
    In these analyses non-spun-up black holes have non-zero spins, as they do in Section~\ref{sec:nonzero_spins}.
    The posteriors on mixing fractions $\lambda$ are not meaningfully different to those seen in Fig.~\ref{fig:nz_lambda_corner}, where all subpopualtions have identical mass ratio and tilt distributions.
    }
    \label{fig:q_tilt_subpops}
\end{figure}

To address this concern, we revisit the model in Section~\ref{sec:nonzero_spins}, now allowing the non-spun-up, singly-spun-up and both-spun-up subpopulations to have non-identically parameterized mass ratio and spin tilt distributions.
To simplify these checks, we assume truncated Gaussian cosine tilt distributions rather than truncated Gaussian -- uniform mixture models.
In Fig.~\ref{fig:q_tilt_subpops}, we plot the posterior distributions for the hyperparameters governing the mass ratio distributions, cosine tilt distributions, and branching fractions for each subpopulation.
We see that the mass ratio and tilt distributions appear to be broadly consistent between all subpopulations.
Moreover, the branching fractions inferred when allowing each subpopulation to have its own mass ratio and tilt distribution are not meaningfully different to those found in Fig.~\ref{fig:nz_lambda_corner}.
This means that a simple, unmodeled covariance between spin magnitude and tilt or mass ratio is unlikely a driving factor in our results.
We cannot, however, rule out the influence of a more subtle unmodeled correlation between parameters without more extensive testing.

\subsection{Informative events}

\begin{figure}
    \centering
    \includegraphics[width=0.7\textwidth]{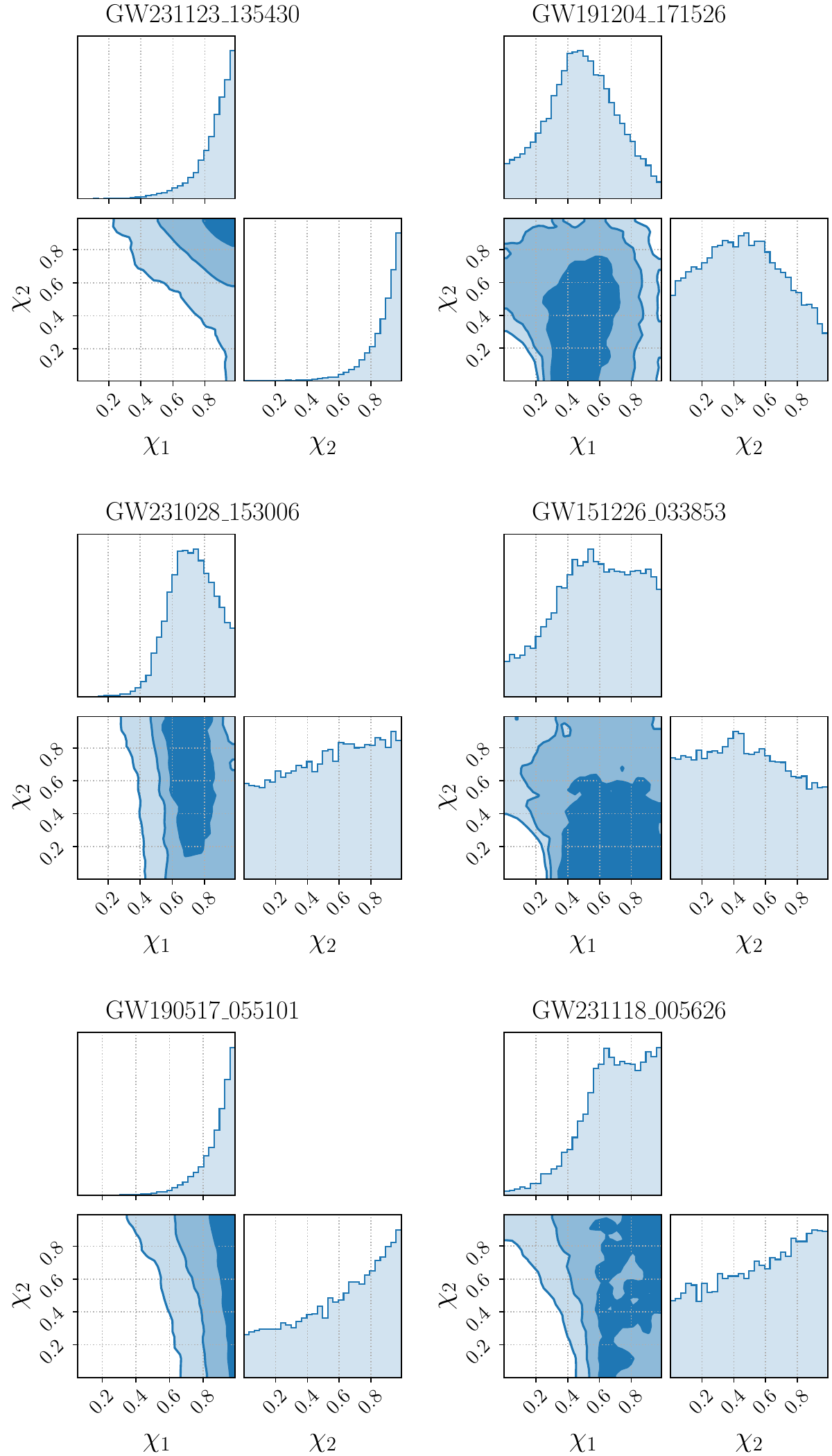}
    \caption{
    Posteriors on component spin magnitudes for the six events with the largest spin-Bayes factors $\ln \mathcal{Z}_\mathrm{spin} - \ln \mathcal{Z}_0$.
    From left-to-right, top-to-bottom, events are ordered from largest ($36$) to smallest ($8$) values of $\ln \mathcal{Z}_\mathrm{spin} - \ln \mathcal{Z}_0$.
    From darkest to lightest, the shaded regions in the two-dimensional contour plots give the 50\%, 90\% and 99\% credible regions.
    }
    \label{fig:spin_posts}
\end{figure}

Here, we investigate the events that we measure most confidently to have non-zero spins.
Namely, in Fig.~\ref{fig:spin_posts} we plot posteriors for component spin magnitudes $\chi_1$ and $\chi_2$ for the six events with the largest spin-Bayes factors ($\ln \mathcal{Z}_\mathrm{spin} - \ln \mathcal{Z}_0 \gtrsim 8$; see Fig.~\ref{fig:spin_evidence} and surrounding text).

A number of these events appear consistent with $\chi_1=0$ \textit{or} $\chi_2=0$ in their marginal distributions, but all strongly rule out the case that both $\chi_1=0$ \textit{and} $\chi_2=0$ in their two-dimensional posteriors.
Furthermore, while all of these events (with the exception of GW231123\_135430) appear marginally consistent with $\chi_2=0$, we see a consistent pattern of the posteriors subtly peaking away from zero (with the exception, perhaps, of GW151226\_033853).
This seems to tell a similar story to Fig.~\ref{fig:spin_evidence}: a small number of events with mostly weak preferences for both components spinning are driving a relatively strong preference for both components spinning across the population.

We see a number of events which have peak posterior support as $\chi_1 \rightarrow 1$ and $\chi_2 \rightarrow 1$ -- namely, GW231123\_135430, GW190517\_055101, GW231118\_005626, and, to a lesser extent GW231028\_153006.
These signals likely contribute much of the evidence for the high-spin subpopulation inferred in Section~\ref{sec:nonzero_spins}.

\bibliographystyle{aasjournal}
\bibliography{refs}

\end{document}